\documentclass[12pt]{article}
\usepackage{epsfig}

\def\fun#1#2{\lower3.6pt\vbox{\baselineskip0pt\lineskip.9pt
  \ialign{$\mathsurround=0pt#1\hfil##\hfil$\crcr#2\crcr\sim\crcr}}}

\newcommand{\dd}{\mbox{d}}

\newcommand{\vecc}[1]{\mbox{\boldmath $#1$}}

\newcommand{\be}{\begin{equation}}
\newcommand{\ee}{\end{equation}}
\newcommand{\ba}{\begin{eqnarray}}
\newcommand{\ea}{\end{eqnarray}}

\title{ Polarized triplet production by circularly polarized photons }

\author{
V.V. Bytev$^{a}$,
E.A. Kuraev$^{a}$
M.V. Galynskii$^{b}$,
A.P. Potylitsyn$^{c}$
\vspace{4mm}
\\
\small  $^a$ Joint Institute for Nuclear Research, Dubna, 141980 Russia \\
\small  $^b$Institute of Physics NAS, Minsk, 220072 Belarus \\
\small  $^c$ Tomsk Polytechnic University, Tomsk, 634050 Russia
}

\date{\today}
\begin{document}
\maketitle

\begin{abstract}
A process of the pair production by a circularly polarized photon
in the field of unpolarized atomic electron has been considered in the
Weizaecker-Williams approximation. The degree of longitudinal polarization
of positron and electron has been calculated. An exclusive cross-section
as well as a spectral distribution are obtained. We estimate the accuracy
of our calculations at the level of a few percent. We show the identity
of the positron polarization for considered process and for process of
pair production in the screened Coulomb field of nucleus.
\end{abstract}

\section*{Introduction}
In this paper we consider the process $\gamma(k)+e^{-}(p)
\to e^{-}(q_{-})+e^{+}(q_{+}) + e^{-}(p^{ '})$ in the high-energy limit
when polarization states of initial photon and production particles
$e^+,\, e^-$ are helical.

\begin{figure}[hb]
\centerline{
\leavevmode \epsfxsize=0.30\textwidth \epsfbox{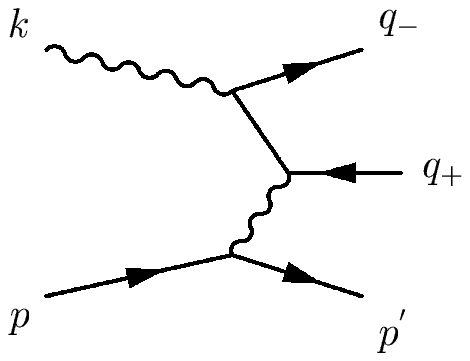} \hspace{1cm}
\leavevmode \epsfxsize=0.30\textwidth \epsfbox{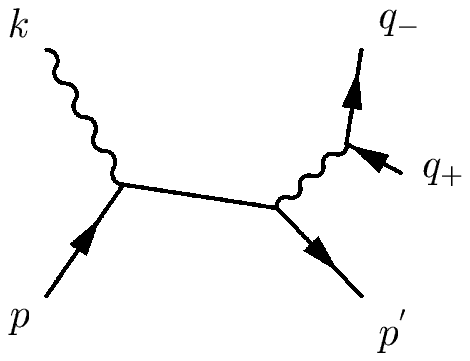}
}
\caption{ Types of relevant Feynman diagrams (left) and irrelevant Feynman diagrams (right)}
\label{fig1}
\end{figure}

The differential cross-section of electron-positron pair photoproduction
on a free electron in the Born approximation is described by eight
Feynman diagrams (FD) \cite{AB} which shown in Fig.\ref{fig1},left.
In the high-energy limit (for photon energy $54$ MeV at Lab system)
only subset of two Bethe-Heitler (BH) diagrams (see Fig.\ref{fig1},left)
are relevant at the level of accuracy at $\frac{m^2}{s}\sim 10^{-2}$,
$m$ is electron mass and $s$ is $s=2kp=2m\omega$, where $\omega$ is the photon
energy at the laboratory frame, whereas the contributions we examine
is order of $\frac{s}{m^2}\sim 200\gg 1$. Non-logarithmic terms we
also don't consider because of their smallness $\sim 10^{-1}$ in
comparison with logarithmic terms $2\ln\frac{s}{m^2}\sim10$.
Further calculations will be performed in Weizaecker-Williams (WW)
approximation \cite{BFKK}.

The differential cross-section of triplet photoproduction
in the Born approximation for the unpolarized case was calculated
numerically in \cite{KOP,Mork} by using Monte Carlo simulation of all
8 FD contribution. The closed analytical expression  is very
cumbersome and was first obtained in a complete form in works \cite{Haug}.
A detailed analysis of the expressions of ~Haug's work reveals that the
interference terms of the BH matrix elements with the other three
gauge-invariant subsets (which take into account the bremsstrahlung mechanism
of pair creation and Fermi statistics for fermions) turn out to be of
the order of some percent for $s>50-60\, m^2$.

The differential cross-section for electron-positron pair production by
linearly polarized photons was derived in a series of papers
\cite{Bold1,Vinokurov,Aku} (see also \cite{Bold2} and
references therein). In Ref.~\cite{Endo} was performed a Monte Carlo
simulation of the process under consideration, in which all eight
lowest order diagrams can be numerically treated without approximation.
There it was shown that one might consider only the two leading graphs
in a wide range of photon energies from $50$ to $550$ MeV. Note that this
observation was made earlier for the unpolarized case in works of ~Kopylov
et al. ~\cite{KOP} and ~Haug ~\cite{Haug} (who presented his results in
explicit analytical form).

The process of the polarized pair production by a polarized photon
in the screened Coulomb field of nucleus have been considered in the high
energy limit in the works \cite{Olsen,Bayer}. Here the degree of
longitudinal polarization of electron has been calculated.

From the paper ~\cite{KOP,Haug} it follows
that: 1. the contribution of FD (see Fig.\ref{fig1},right) as well as interference
it's amplitude with amplitude of (see Fig.\ref{fig1},left) can be neglected
within accuracy $3\%$ compared with contribution of (see Fig.\ref{fig1},left)
already starting from photon energies $\omega>30$ MeV in laboratory frame;
2. One can use the asymptotic formulae at very high photon energies of
contribution of ~Bethe-~Haitler FD (see Fig.\ref{fig1},left) within accuracy
$5\%$ starting from energies of order $\omega>100$ MeV.
Taking into account that unpolarized cross-section dominates the ones
depending on particle polarization we estimate accuracy of WW
(logarithmical approximation) on the level of $10\%$.

Our paper organized as follows. Using the Sudakov technique we calculate the
differential cross-section and the degree of longitudinal polarization of
electron and positron in the process of pair production by circularly
polarized photon on electron. In conclusion we discuss the different
schemes of production of longitudinally polarized positrons in experimental
set-ups. The method described here is one of the most perspective one.
It's our motivation for investigation. In Appendix we give in Born
approximation the differential cross-section of triplet production
process for the case when polarization states of all particles are
helical  by using of crossing-transformation of corresponding differential
cross-section for M\"oller bremsstrahlung process in the
ultrarelativistic (massless) limit. Corresponding formulae for the degree of
longitudinal polarization of positron in terms of kinematical invariant and helicity
of initial photon is given for a case of high energy large angle scattering.
\section*{The cross-section of the process} \label{sec:2}
The cross-section of the process is proved in the next form:
\ba
\dd\sigma=\frac{(4\pi\alpha)^3}{4(2\pi)^5s}\Sigma M^2\frac{\dd^3q_{-}}
{2\varepsilon_{-}} \frac{\dd^3q_{+}}{2\varepsilon_{+}}\frac{\dd^3p^{'}}
{2\varepsilon^{'}}
\times \, \delta^4(p+k-q_{+}-q_{-}-p^{'})\, ,
\ea
where $p^{'}, \, \varepsilon^{'}$ is four-momentum and energy of scattered
electron; $q_{\pm}$ and $\varepsilon_{\pm}$ are four-momentum and energy of
electron and positron correspondingly. We accept here the Sudakov picture
\cite{BFKK} of the peripherical process.
Let us determine the basic vectors in Sudakov representation of kinematics:
vector of momentum transfer $q=p-p^{'}$ and the light-like vector $\tilde{p}=
p-k\, \frac{m^2}{s},  \, \tilde{p}^2=0$ ($k$ is 4-momentum of impact photon).
The Sudakov representation of four-momenta are the next:
\ba
q=\alpha\tilde{p}+\beta k+q_{\bot},\qquad
q_\pm=\alpha_\pm\tilde{p}+\beta_\pm k+q_{\bot\pm}, \qquad
a_\bot k=a_\bot \tilde{p}=0.\nonumber
\ea
The quantities $\beta_{\pm}$ can be interpreted as the energy fraction
of pair components of photon energy, $\beta_{+}+\beta_-=1$. The conservation
law of transversal momenta is $q_\bot=q_{\bot+}+q_{\bot-}$. Momentum
components along $\tilde{p}$ is small. Using the mass-shell conditions
of pair components and the scattered electron leads to:
\ba
s\alpha_\pm=\frac{\rho_\pm}{\beta_\pm},\qquad
s\beta(1-\alpha)=-\vecc{q}^{2}-m^2\alpha,\qquad
\rho_\pm=\vecc{q}_\pm^{2}+m^2,
\ea
where we use the two-dimensional Euclidean vectors $q_\bot^2=-\vecc{q}^{2},
\; q_{\bot\pm}^{2}=-\vecc{q}_\pm^{2}$. The Sudakov parameter $\alpha$
can be related with the invariant mass of created pair
\ba
s_1=(q_{+} + q_{-})^2, \, s \alpha=s_1+{\vecc{q}}^{2}\,.
\ea
Using conditions of mass-shell and smallness of Sudakov parameter
$\alpha$ the momentum transfer to the target square can be written
in the next form:
\ba
q^2=-\vecc{q}^{2}-m^2 \, \biggl(\frac{s_1}{s}\biggr)^2 \,.
\ea
We see that $q^2<0$, is negative and has non-zero magnitude.
Using the Sudakov representation of momentum phase volume of the particle
\ba
\dd^4q=\frac{s}{2}\, \dd\alpha \, \dd\beta \,\dd^2q_\bot
\ea
we perform the phase volume of the final state:
\ba
\dd\Gamma=\frac{\dd^3q_{-}}{2\varepsilon_{-}}
\frac{\dd^3q_{+}}{2\varepsilon_{+}}\frac{\dd^3p^{'}}
{2\varepsilon^{'}}\; \delta^4(p+k-q_{+}-q_{-}-p^{'})
=\frac{1}{2s}\frac{\dd\beta_-}{\beta_-\beta_+}\dd^2q_-\dd^2q.
\label{obem}
\ea
Let us now consider the matrix element:
\ba
M=\bar{u}(p^{'})\gamma^\mu u(p)\frac{1}{q^2}
e_\rho(k)\bar{u}(q_-)O^{\rho\nu}v(q_+)g_{\mu\nu}. \nonumber
\ea
By using the Gribov's decomposition of metric tensor and omitting
the terms of order $m^2/s$ compared of the terms of order of unity
\ba
g_{\mu\nu}=g_{\mu\nu}^\bot+ \frac{2}{s}\biggl(\tilde{p}_\mu k_{\nu}+
\tilde{p}_\nu k_{\mu}\biggr) \approx\frac{2}{s}\tilde{p}_\nu k_{\mu}\,,
\ea
one can obtain the matrix element in the next form:
\ba
M=\frac{-2s N_p}{\vecc{q}^{2}+m^2(\frac{s_1}{s})^2}
e_\rho(k)\,\bar{u}(q_-)V^{\rho}v(q_+)\; , \\
V^\rho=\frac{1}{s}O^{\rho\nu}\tilde{p}_{\nu},\qquad
N_p=\frac{1}{s}\bar{u}(p^{'})\hat{k}u(p),\qquad |N_p|=1.  \nonumber
\ea
Using the mass-shell conditions we can express the matrix
four-vector $V_\rho$ as:
\ba
V_\rho=\beta_{+}\beta_{-}\bigl(\frac{1}{\rho_-}
-\frac{1}{\rho_+}\bigr)\gamma_\rho
+\frac{\beta_{+}}{\rho_+}\hat{\tilde{p}}\hat{q}\gamma_\rho
-\frac{\beta_{-}}{\rho_-}\gamma_\rho\hat{q}\hat{\tilde{p}}.
\ea
One can see that the quantity $V_\rho$ is proportional to $|q_\bot|$
at small $|q_\bot|$. Weizaecker-Williams approximation corresponds
to the logarithmical enhancement factor:
\ba
\int\frac{\dd\vecc{q}^{2}\vecc{q}^{2}}{(\vecc{q}^{2}+m^2\bigl
(\frac{s_1}{s}\bigr)^2)^2} \approx 2\ln\frac{s}{s_1}\,.
\ea
The polarization matrix of density $\tau^{\delta_{\pm}}_{\pm}$ for
$e^{\pm}$ particles have the next form:
\ba
\tau^{\delta_{\pm}}_{\pm} = \frac{1}{2} (\hat q_{\pm} \mp m)
(1 - \delta_{\pm} \gamma^{5} \hat s_{\pm} ) ,\qquad
\gamma^{5} \hat s_{\pm} \tau^{\delta_{\pm}}_{\pm}= \delta_{\pm}
\tau^{\delta_{\pm}}_{\pm}.
\ea
We will express the particle spin vectors $s_{\pm}$
in terms of the 4-momenta $q_{\pm}$ and $\tilde{p}$,
\ba
s_{\pm}=\frac{\mp q_{\pm}}{m}\pm \frac{2 m \tilde{p}}{s \beta_{\pm}}, \qquad
s_{\pm} q_{\pm}=0 , \, s_{\pm}^2=-1\,.
\ea
The circular-polarization vector $e^{\lambda}$ of a photon with 4-momentum
$k$ is conveniently defined by using the 4-vectors $ q_{+},\, q_{-}$ and
$k$ ~\cite{GAL89}:
\ba
e^{\lambda}_{\mu} = { (q_- k)(q_+)_{\mu} -(q_+k)(q_-)_{\mu}
+ i \lambda \; \varepsilon_{\mu\nu\rho\sigma}q_+^\nu q_-^\rho k^\sigma
\over  \sqrt{2\,z} }\,  , \nonumber\\
z = -\,( (q_-k) q_+ - (q_+k) q_-)^2={\rho^2(\rho-m^2)\over 4 \beta_+^2
\beta_-^2}\,.\nonumber
\ea
Performing the angular averaging on $\dd^2q$ and extracting the WW-factor
we write down the differential cross-sections in the form (in WW-approximation
we can put $\rho_+=\rho_-=\rho$):
\ba
\label{allpol}
\frac{\dd\sigma}{\dd\vecc{q}^{2}_-\dd\beta_-}=\frac{2\alpha^3}{\rho^2}
\ln\frac{s}{m^2} \biggl[1-2\beta_+\beta_-
+\frac{4m^2}{\rho^2}(\rho-m^2)\beta_+\beta_- \nonumber \\
+\lambda\, \delta_+\biggr(\beta_+-\beta_-
+\frac{4m^2}{\rho^2}(\rho-m^2)\beta_-\biggl)\,  
+\lambda\, \delta_-\biggr(\beta_+ -\beta_-
-\frac{4m^2}{\rho^2}(\rho-m^2)\beta_+\biggl) \nonumber  \\
+\delta_+\delta_-(-6\beta_+\beta_- +{\rho-m^2 \over m^2}
-\frac{4m^2}{\rho^2}(\rho-m^2)(1-3\beta_+\beta_-))\biggr]. \nonumber
\ea
In the case when initial electron is being polarized formulae given above
isn't changed. The corresponding effects are of order $m^2/s$.
Performing the summation over the polarization of electron in (\ref{allpol})
we will get the differential cross-sections
for creating polarized positron:
\ba
\frac{\dd\sigma^+}{\dd\vecc{q}^2_-\dd\beta_-}&=&
\frac{4\alpha^3}{\rho^2} \ln \frac{s}{m^2}
\biggl[1-2\beta_+\beta_-
+ \frac{4m^2}{\rho^2}(\rho-m^2)\beta_+\beta_{-} \nonumber\\
 &+& \lambda\delta_{+}\biggr(\beta_+-\beta_-
+\frac{4m^2}{\rho^2}(\rho-m^2)\beta_-\biggl)\biggr].   
\label{polps}
\ea
and analogously expression for cross-section of creating polarized electron:
\ba
\frac{\dd\sigma^-}{\dd\vecc{q}^2_-\dd\beta_-}&=&
\frac{4\alpha^3}{\rho^2} \ln \frac{s}{m^2} \biggl[1-2\beta_+\beta_-
+ \frac{4m^2}{\rho^2}(\rho-m^2)\beta_+\beta_{-}\nonumber\\
&+& \lambda\delta_{-}\biggr(\beta_+ - \beta_-
-\frac{4m^2}{\rho^2}(\rho-m^2)\beta_+\biggl)\biggr].   
\label{polel}
\ea
>From expressions (\ref{polps}) and (\ref{polel}) we will have the degree
of longitudinal polarization of electron (positron) when polarization
of positron (electron) no registered:
\ba
\label{21}
\delta_{\pm}^f=\lambda \frac{\beta_+-\beta_-
\pm\frac{4m^2}{\rho^2}(\rho-m^2)\beta_{\mp}}
{1-2\beta_+\beta_-
+ \frac{4m^2}{\rho^2}(\rho-m^2)\beta_+\beta_{-}} \,.
\ea
The result (\ref{21}) is in agreement with one given in more general form in
\cite{Bayer} (see formulae (19.8)).
Performing the integration over transversal momenta of pair we obtain for
the spectral distribution:
\ba
\frac{\dd\sigma^+}{\dd\beta_-}=4\alpha r^2_0\ln\frac{s}{m^2}\biggl[
1-\frac{4}{3}\beta_+\beta_-+\lambda\delta_{+}\biggl(1-\frac{4}{3}\beta_-\biggr)\biggr].
\ea
Here we use in WW-approximation $|\vecc{q_+}|=|\vecc{q_-}|, \, \rho=\rho_+=
\rho_-=\vecc{q^2_-}+m^2$, $\lambda$ is the degree of circular polarization
if the initial photon. The degree of longitudinal polarization of
created $e^{\pm}$ particles have a form:
\ba
\label{22}
\delta^f_{\pm}=\lambda\frac{1-\frac{4}{3}\beta_\mp}
{1-\frac{4}{3}\beta_+\beta_-}\, .
\ea
We see from (\ref{22}) that in the limit $\beta_+\to 1 (\beta_-\to 1)$
the degree of longitudinal polarization of positron (electron)
equal to degree of circularly polarization of initial photon.

Let's compare the result obtained with calculations of longitudinal positron
polarization from pair production process in the screened Coulomb field of
a nucleus \cite{Olsen} with the same accuracy as before we may write (see \cite{Pot}):
\ba
\delta^f_{+} \approx \frac{\frac{4}{3}\beta_{+}-\frac{1}{3}} {(\beta^2_{+}+
\beta^2_{-})+ \frac{2}{3}\beta_{+}\beta_{-}}\lambda .
\ea
Having in mind that $\beta_{-} = 1-\beta_{+}$,  one may obtain from
(\ref{22}) the same result.

\begin{figure}[h]
\vspace{-0.5cm}
\centerline{
\leavevmode \epsfxsize=0.55\textwidth \epsfbox{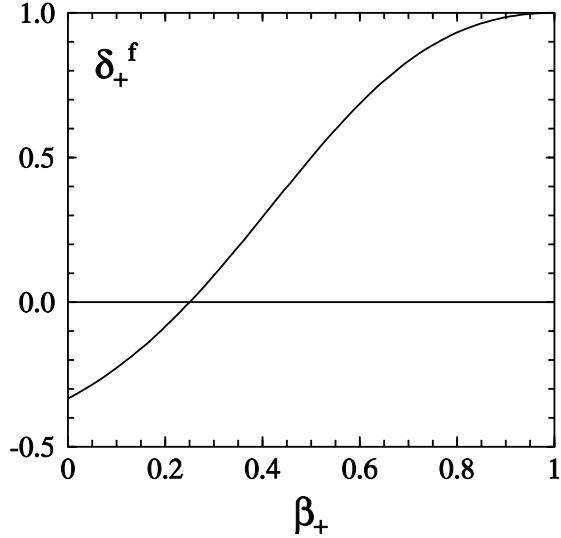}
}
\vspace{-0.5cm}
\caption{ The degree of longitudinal polarization of
positron $\delta_+^f$ (\ref{22}) versus of $\beta_+$ }
\label{fig3}
\end{figure}
\vspace{-0.2cm}

\section*{Conclusions}
A few schemes to create polarized positrons for electron-positron
colliders were proposed \cite{Clend}. Longitudinally polarized positrons
are produced in the pair production process by circularly polarized
photon with energy $>$ 10 MeV. A circularly polarized $\gamma$-beam
is formed due to undulator radiation of electrons with energy
$E \sim 10^2$ ~GeV in a helical undulator \cite{Balakin} or due to Compton
backscattering process of the circularly polarized laser photons
on the electron beam with energy $>$ 1 ~GeV \cite{Dobas}. In both cases
there was considered the process of pair production in the screened
coulomb field of nucleus only as a source of polarized positrons.
The process of triplet production was considering as background
process \cite{Pot}.
The obtained results allow to develop a correct Monte Carlo code in
order to receive the positron polarization for the real experimental
situation taking into account both processes of pair production in
amorphous target \cite{Kotserog}.

In the recent experiment \cite{Sakai} authors measured the circular polarization
of $\gamma$-quanta with energy $>$ 20 MeV using the magnetized iron polarimeter.
The thickness of iron substance was too big (7 cm), so the correct estimation of the
analyzing power may be done by a Monte-Carlo technique. In this case the polarization
characteristics of all reaction particles should be took into account and our
results cover the existing lack.
\section*{Appendix}
Differential cross section for triplet production process one can easily
get with help of crossing transformation of expression for square
amplitudes of M\"oller bremsstrahlung process
\ba
e^{-}(p_{1})+e^{-}(p_{2}) \to e^{-}(p_{3}) + e^{-}(p_{4}) + \gamma(k) \; ,
\label{Meller}
\ea
which corresponds eight Feynman diagrams. In papers \cite{GAL89}  was
calculated the differential cross section for this reaction in the case
when all fermions were massless ($p_{i}^{2} = 0$, where $i = 1, 2, 3, 4 $)
with taking into account polarization of initial electrons and emitted photon.
Let us introduce invariant variables :
\ba
s_1 = ( p_{1} + p_{2} )^{2} \; , \; t_1 = ( p_{1} - p_{3} )^{2} \; ,
\; u_1 = ( p_{1} - p_{4} )^{2} \; ,\\
\label{inv1}
s_2 = ( p_{3} + p_{4} )^{2} \; , \; t_2 = ( p_{2} - p_{4} )^{2} \; ,
\; u_2 = ( p_{2} - p_{3} )^{2} \; ,\nonumber
\ea
and helicities $\delta_1$ and $\delta_2$ for initial electrons
with momentum $p_1$ and $p_2$ respectively, $\lambda$ for helicity
of emitted photon. The differential cross section of reaction
(\ref{Meller})
in the case of helically polarized initial electrons and photon has the
next form \cite{GAL89}:
\ba
\label{sigma}
d \sigma_{M} = { \alpha^{3} \over 2 \pi^{2} s } \, A_{M} \, W_{M} \,
d \Gamma_M  , \qquad
A_{M} =\frac{ A_{MB}}{ t_1  t_2  u_1  u_2 } \; ,
\ea
\ba
d \Gamma_M =
{d^{3} \vecc p_{3} \over 2 p_{30}} \; {d^{3} \vecc p_{4} \over 2 p_{40}}
\; {d^{3} \vecc k \over 2 \; \omega} \,
\delta^{4} ( p_{1} + p_{2} -p_{3} - p_{4} - k )\;.
\nonumber
\ea
Expressions for $A_{MB}$ and $W_{M}$ have forms \cite{GAL89}:
\ba
A_{MB}& =&  ( 1 + \delta_1 \delta_2 ) \; [ ( 1 + \delta_1 \lambda )
 s_1 s_2 s_2^{2} + ( 1 - \delta_1 \lambda ) s_1 s_2 s_1^{2} ] +\nonumber\\
&+& ( 1 - \delta_1 \delta_2 ) \; [ ( 1 + \delta_2 \lambda ) \;
( t_1 t_2 t_1^{2} + u_1 u_2 u_1^{2} )+ \nonumber\\
&+& ( 1 - \delta_2 \lambda ) \;
( t_1 t_2 t_2^{2} + u_1 u_2 u_2^{2} ) ] \; ,\label{AMBN}\\
 W_{M}&=& -\, \left ( {p_{1}\over p_{1} k} + {p_{2}\over p_{2} k} -
{p_{3}\over p_{3} k} - {p_{4}\over p_{4} k} \right )^{2} \; .
\label{WMN}
\ea
Photoproduction diagrams of triplet
\ba
\gamma(k) + e^{-}(p) \rightarrow  e^{-}(q_{-}) + e^{+}(q_{+})+e^{-}(p^{'})
\label{triplt}
\ea
are different from ones for M\"oller bremsstrahlung process (\ref{Meller})
with exchange
\ba
p_1 \to p\,, \, p_2 \to -q_+\,, \, p_3 \to p^{'}\,, \, p_4 \to q_- \,,
\label{cross}\\
k \to -k \,, \, \lambda \to -\lambda\,, \, \delta_1 \to \delta_1\,, \,
\delta_2 \to -\delta_+\,\,.\nonumber
\ea
Both processes (\ref{Meller}) and (\ref{triplt}) are two crossing canals
with the same (generalized) reaction. After the replacement (\ref{cross})
of invariant variables (\ref{inv1}) they takes the next form:
\ba
&&s_1 = ( p - q_+)^2 , \;  t_1 = ( p - p{'} )^2 \; ,
\; u_1 = ( p - q_- )^2 \; ,\\
\label{inv2}
&&s_2 = ( p^{'} + q_- )^2  , \, t_2 = ( q_{+} + q_{-} )^2  ,
\, u_2 = ( p^{'} + q_{+} )^2 \, .\nonumber
\ea
Since square amplitude for process (\ref{triplt}) after replacement of
variables (\ref{cross}) is like to one for process (\ref{Meller}),
then one can obtain after initial electron polarization summing
in the reaction (\ref{triplt}) from (\ref{sigma}) the next formulae
for triplet production with taking into account polarization
of initial photon and  scattered positron:
\ba
d \sigma_{tr}& =& { \alpha^{3} \over 2 \pi^{2} s } \, A_{tr} \, W_{tr} \,
d \Gamma  , \,\label{sigmatr}\\
 A_{tr}& =& {A_{MB}^{tr} \over t_1  t_2  u_1  u_2}\,  ,
 A_{MB}^{tr}=A^{0}_{tr}+\delta_+ \lambda \, A^{1}_{tr}\; ,\nonumber\\
A^{0}_{tr}& =& s_1 s_2 ( s_1^{2} + s_2^{2} ) + t_1 t_2 ( t_1^{2} + t_2^{2} ) +
u_1 u_2 ( u_1^{2} + u_2^{2} ) \, ,\nonumber \\
 A^{1}_{tr}&=& - s_1 s_2 ( s_1^{2} - s_2^{2} ) + t_1 t_2
( t_1^{2} - t_2^{2} ) + u_1 u_2 ( u_1^{2} - u_2^{2} )  \; ,\nonumber\\
W_{tr}& =&-\, \left ( { p \over p \,k} + {q_{+}\over q_{+} k} -
{p^{'}\over p^{'} k} - {q_{-}\over q_{-} k} \right )^{2} \; ,
\label{WMN1}
\ea
were $s=(p+k)^2$ and $d \Gamma$ are defined by expression (\ref{obem}).
Then for the degree of longitudinal polarization of produced positron
$\delta_{+}^f=A^{1}_{tr}/A^{0}_{tr}$
we will have expression:
\ba
\delta_{+}^f=
\lambda { - s_1 s_2 ( s_1^{2} - s_2^{2} ) + t_1 t_2
( t_1^{2} - t_2^{2} ) + u_1 u_2 ( u_1^{2} - u_2^{2} )
\over s_1 s_2 ( s_1^{2} + s_2^{2} ) + t_1 t_2 ( t_1^{2} + t_2^{2} ) +
u_1 u_2 ( u_1^2 + u_2^2 )
}. \nonumber
\ea
This formulae is valid only in the case when
invariant mass square of any particle pair is large with comparison to
electron mass square.


\begin{thebibliography}{99}
\bibitem{AB}
A.\,I. Akhiezer and V.\,B. Berestetskiy, Quantum Electrodinamics,
Moscow, Science, 1981.
\bibitem{BFKK}
V.\,N. Baier, V.\,S. Fadin, V.\,A. Khoze and E.\,A. Kuraev,
Phys.~Rep. {\bf 78} 293 (1981).
\bibitem{KOP} G. Kopylov, L. Kuliukina, and I. Polubarinov,
Zh.~Exp.~Teor.~Fiz. {\bf 46, N 5}, 1715 (1964).
\bibitem{Mork} K.\,J.Mork, Phys.~Rev. {\bf 160}, 1065 (1967).
\bibitem{Haug} E. Haug, Z.~Naturforschung {\bf A30} 1099 (1975);
Phys.~Rev. {\bf D31} 2120 (1985); {\bf D32} 1594 (1985).
\bibitem{Bold1}  V.\,F. Boldyshev and Yu.\,P.Peresun'ko, Yad.~Fiz.,
{\bf 13}, 588 (1971).
\bibitem{Vinokurov} E.\,A.Vinokurov and E.\,A.Kuraev, JETP {\bf 36}, 602 (1973).
\bibitem{Aku} I.\,V.Akushevich, H.Anlauf, E.\,A.Kuraev et al.
Phys.~Rev., {\bf A61}, 032703 (2000).
\bibitem{Bold2} V.\,F.Boldyshev, E.\,A.Vinokurov, N.\,P.Merenkov et al.,
Fiz.~Elem.~Chastits At.~Yadra, {\bf 25} 696 (1994).
\bibitem{Endo} I.Endo and T.Kobayashi, Nucl.~Inst.~Meth., {\bf A328} 517 (1993).
\bibitem{Olsen} H.Olsen, L.\,C.Maximon. Phys.~Rev. {\bf 114} 887 (1959).
\bibitem{Bayer} V. Baier, V. Fadin, V. Katkov, Emission of
relativistic electrons. M. Atomizdat, 1973.
\bibitem{GAL89} M.\,V. Galynskii, L.\,F. Zhirkov, S.\,M. Sikach et al.,
Zh.~Exp.~Teor.~Fiz. {\bf 95}, 1921 (1989);
M.\,V. Galynskii and S.\,M. Sikach, Fiz.~Elem.~Chastits At.~Yadra
{\bf 29}, 1133 (1998).
\bibitem{Pot} A.\,P. Potylitsyn. Nucl.~Instrum.~Meth. {\bf A398} 395 (1997).
\bibitem{Clend} J. Clendenin. Recent Advances in Electron and Positron
Sources. SLAC-PUB-8465 (2000).
\bibitem{Balakin} V.Balakin, A.\,A. Mikhailichenko. Preprint INP 79-85,
Novosibirsk, 1979.
\bibitem{Dobas} K.Dobashi, et al., Nucl.~Instrum.~Meth. {\bf A437},
169 (1999).
\bibitem{Kotserog} T. Kotseroglon, V. Bharadwaj, J.\,E. Clendenin et al.
Proceedings of the 1999 PAC, New York, p. 3450 (1999).
\bibitem{Sakai} I. Sakai, T. Hirose, K. Dobaski et al.
{\bf 21} ICFA Workshop, Stony Brook, 2001;
\end{thebibliography}
\end{document}